\documentclass[english,english,noshowpacs,amssymb,aps,nofootinbib]{revtex4-2}
\usepackage{graphicx}
\usepackage{amsmath}

\newcommand{\sech}{\mathrm{sech}}
\newcommand{\be}{\begin{equation}}
\newcommand{\ee}{\end{equation}}
\newcommand{\bea}{\begin{eqnarray}}
\newcommand{\eea}{\end{eqnarray}}
\newcommand{\ket}[1]{\left|#1\right\rangle}
\newcommand{\expected}[1]{\left\langle#1\right\rangle}
\newcommand{\var}[1]{\expected{\Delta^2#1}}

\begin{document}

\title[Fundamental sensitivity bounds for optical resonance sensors]{Fundamental Sensitivity Bounds for Quantum Enhanced Optical Resonance Sensors Based on Transmission and Phase Estimation}

\author{Mohammadjavad~Dowran$^{1,2}$, Timothy~S.~Woodworth$^{1,2}$, Ashok~Kumar$^{1,3}$, and Alberto~M.~Marino$^{1,2}$}
\address{$^1$Homer L. Dodge Department of Physics and Astronomy, The University of Oklahoma, Norman, Oklahoma 73019, USA}
\address{$^2$Center for Quantum Research and Technology (CQRT), The University of Oklahoma, Norman, Oklahoma 73019, USA}
\address{$^3$Indian Institute of Space Science and Technology, Thiruvananthapuram, Kerala 695 547, India.}

\begin{abstract}
Quantum states of light can enable sensing configurations with sensitivities beyond the shot-noise limit (SNL). In order to better take advantage of available quantum resources and obtain the maximum possible sensitivity, it is necessary to determine fundamental sensitivity limits for different possible configurations for a given sensing system. Here, due to their wide applicability, we focus on optical resonance sensors, which detect a change in a parameter of interest through a resonance shift. We compare their fundamental sensitivity limits set by the quantum Cram\'er-Rao bound (QCRB) based on the estimation of changes in transmission or phase of a probing bright two-mode squeezed state (bTMSS) of light. We show that the fundamental sensitivity results from an interplay between the QCRB and the transfer function of the system.  As a result, for a resonance  sensor with a Lorentzian lineshape a phase-based scheme outperforms a transmission-based one for most of the parameter space; however, this is not the case for lineshapes with steeper slopes, such as higher order Butterworth lineshapes. Furthermore, such an interplay results in conditions under which the phase-based scheme provides a higher sensitivity than the transmission-based one but a smaller degree of quantum enhancement. We also study the effect of losses external to the sensor on the degree of quantum enhancement and show that for certain conditions probing with a classical state can provide a higher sensitivity than probing with a bTMSS. Finally, we discuss detection schemes, namely optimized intensity-difference and optimized homodyne detection, that can achieve the fundamental sensitivity limits even in the presence of external losses.
\end{abstract}

\maketitle


\section{Introduction}\label{intro}
The field of quantum sensing seeks to take advantage of unique quantum properties to enhance the precision of sensing techniques and devices beyond the fundamental classical bound given by the shot noise limit (SNL)~\cite{RevModQuantumSensing}.
Such a quantum enhancement can be achieved through the use of quantum states with reduced noise properties, such as squeezed states, to reduce the uncertainty in the estimation of a parameter of interest~\cite{caves1981quantum,xiao1987precision,giovannetti2011advances,lawrie2019quantum}.
This approach can enable quantum based sensitivity enhancements when the detection techniques and sensing devices are operating at the SNL and are able to preserve the quantum properties of the quantum states. Such approaches have already been implemented in LIGO through the coupling of a vacuum squeezed state into one of the input ports of the interferometer to enhance its sensitivity~\cite{abadie2011gravitational}.
Applications of optical quantum enhanced devices range from quantum enhanced interferometry~\cite{grangier1987squeezed} to quantum sensing~\cite{holtfrerich2016toward,dowran2018quantum,xu2019sensing,pooser2016plasmonic} to quantum imaging~\cite{treps2003quantum,brida2010experimental} to biological sensing~\cite{taylor2013biological,taylor2016quantum}.

Sensors based on optical readout techniques detect changes in the probing electromagnetic field to estimate a change in the parameter of interest~\cite{yesilkoy2019optical}. Here, we focus on optical resonance sensors, such as optical cavities~\cite{zhi2017single}, whispering-gallery mode sensors~\cite{foreman2015whispering,wade2016applications}, photonic crystal sensors~\cite{pitruzzello2018photonic}, and plasmonic sensors~\cite{homola2003present,deng2017phase}, due to their wide applicability as  optical readout label-free sensors.  These sensors exhibit a resonance in their response that can be used to estimate changes in a physical quantity of interest, such as temperature, pressure, force, index of refraction, etc., through measurements of changes in either the transmission or phase of the probing light. Given that the sensitivity of optical resonance sensors has already reached the SNL when probed with classical states of light~\cite{subramanian2018label}, the use of quantum states is necessary to enable a quantum-based enhancement to obtain a sensitivity beyond this limit. For example, the use of a bright two-mode squeezed state (bTMSS) has been shown to enable quantum enhancements in the estimation of phase~\cite{anderson2017optimal,anderson2017phase} and transmission~\cite{woodworth2020transmission}.
In this theoretical study, we compare the sensitivity limits given by the quantum Cram\'er-Rao bound (QCRB)~\cite{Helstrom1976,Holevo1982,PARIS2009,Tan2019} for optical resonance sensors based on the estimation of changes in transmission or phase of the probing light when probed with either a coherent state or a bTMSS. Given that the QCRB provides the fundamental theoretical bound for sensitivity when proving with a given state of light, it allows us to perform an absolute comparison between an approach based on the estimation of transmission, transmission-based scheme, or the estimation of phase, phase-based scheme.

The paper is outlined as follows: In section~\ref{ResonanceSensing}, we define the sensitivity of optical resonance sensors based on the estimation of changes in the phase or transmission of the  probing light. Then, in section~\ref{QCRB}, we present the QCRB for transmission and phase estimation with bTMSS and coherent states of light in the presence of losses external to the resonance sensor. In section~\ref{comparison}, through the definition of a general transfer function for the sensor, we compare the fundamental sensitivity limits of resonance sensors  for the transmission- and phase-based schemes obtained for estimations at the QCRB. Finally, we discuss the effect of optical losses external to the sensor on the degree of quantum enhancement in section~\ref{Losses} and point to the measurement strategies that can saturate the QCRB for the transmission- and phase-based schemes in section~\ref{SensingSchemes}.

\section{Optical Resonance Sensing Schemes}\label{ResonanceSensing}

\begin{figure}[htb]
  \centering
  \includegraphics{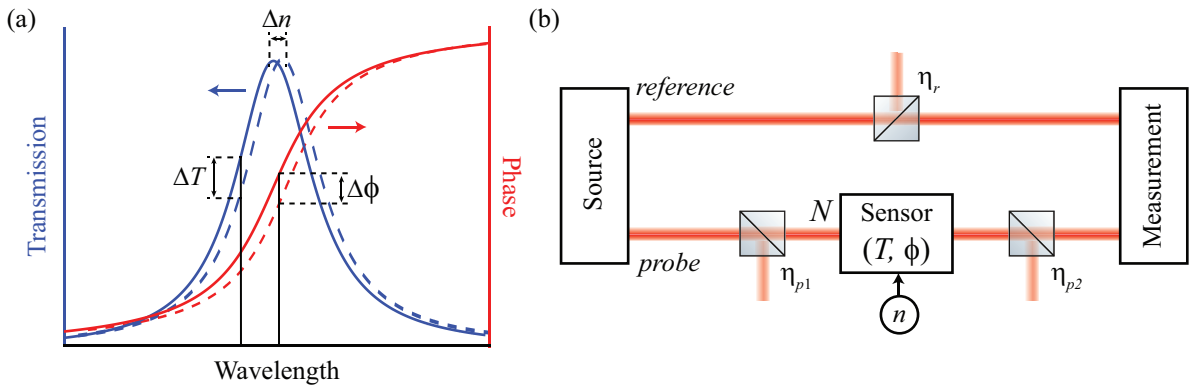}
  \caption{(a) Transmission (blue traces) and phase (red traces) transfer functions for an optical resonance sensor. A change in the physical parameter of interest, $n$, by an amount  $\Delta n$ leads to a shift of the resonance, which results in a corresponding change in transmission, $\Delta T$, and phase, $\Delta \phi$,  of the probing light. These changes in transmission and phase can therefore be used to estimate changes in the physical quantity of interest. Note that the optimal probing wavelength will be different for an optimal transmission or phase change. (b) The sensing cofiguration we consider is composed of a source, either of quantum (bTMSS) or classical (coherent) states of light, a resonance sensor responsive to changes in the physical quantity of interest, $n$, and a detection system to extract information on the change in transmission or phase of the probing light. Optical losses due to imperfections are taken into account with beam splitters with transmission $\eta_{p1}$ before and $\eta_{p2}$ after the resonance sensor in the probe and $\eta_{r}$ in the reference beam path. A fixed number of photons, $N$, is assumed to be incident on the resonance sensors and gives the resources used for the estimation of the quantity of interest.}  \label{Fig:Del_nShift}
\end{figure}

We specialize on resonance sensors that are passive optical devices with a linear response characterized by a transfer function with a resonance consisting of either a dip or a peak in its transmission spectrum, $T(\lambda)$ where $\lambda$ is the wavelength. When such a sensor is probed with light, in addition to a change in transmission, the electromagnetic field undergoes a corresponding change in phase.  This leads to a phase spectrum, $\phi(\lambda)$,  whose relation to the transmission spectrum is governed by the Kramer-Kronig relations~\cite{kop1997kramers,gralak2015phase}.  A change of the external physical quantity of interest, $n$, leads to a change in both the amplitude and phase of the probing light due to a shift of both the transmission and phase spectra\cite{zhou2016performance,kabashin2009phase,lodewijks2012boosting,humer2012phase}, as shown in figure~\ref{Fig:Del_nShift}(a). Thus, the external physical quantity of interest can be measured through two different schemes, one based on transmission estimation, transmission-based scheme, and the other based on phase estimation, phase-based scheme, of the probing light.

To study the behavior and characteristic response of a sensor, we define the sensitivity as the inverse of the uncertainty in the estimation of the physical quantity of interest, $\left(\Delta^2 n\right)$, based on the estimation of a given parameter $X$~\cite{dowran2018quantum,piliarik2009surface}:
\be
\frac{1}{S(n|X)}=\Delta^2 n(X)= \left(\frac{\Delta^2X}{|\partial X/\partial \lambda|^2}\right) \frac{1}{|\partial \lambda/\partial n|^2}.
      \label{Eq:sensitivity}
\ee
For an optical resonance sensor, $X$ represents either the transmission, $T$, or phase, $\phi$, of the probing light. Based on this definition, a better estimation of the physical quantity of interest, i.e. smaller uncertainty in the estimation, leads to a better sensitivity characterized by larger values of $S$. The first term in parenthesis on the right hand side of equation~\eqref{Eq:sensitivity} represents the inverse of the sensitivity in estimating the shift in the resonant wavelength, that is $\left[S(\lambda|X)\right]^{-1}=\Delta^2\lambda(X)=\Delta^2X\big/|\partial X/\partial \lambda|^2$, and depends on the uncertainty in the estimation of parameter $X$, $\Delta^2X$, and its rate of change with wavelength.
The second term on the right hand side of equation~\eqref{Eq:sensitivity}, $|\partial \lambda/\partial n|^{-2}$, depends on the rate of change of the resonance wavelength with respect to the external physical quantity of interest. Given that this second term is a property of the sensor that is common for both sensing schemes, we neglect it in our comparison of the schemes and consider $S(\lambda|X)$ as a measure of the sensitivity of the optical resonance sensor.

The general sensing configuration that we consider is shown in figure~\ref{Fig:Del_nShift}(b).
We compare the use of either a bTMSS for the quantum state or a single mode coherent state, as it gives the optimal classical configuration. For the case of the bTMSS, one mode, denoted by subscript $p$, is used to probe the resonance sensor while the other mode, denoted by subscript $r$, is used as a reference. On the other hand, for the case of a coherent state, we consider only the mode that is used to probe the sensor, as the inclusion of a second mode would only add noise to the estimation and would thus reduce the sensitivity of the sensor~\cite{PARIS2009,Tan2019}.  Furthermore, we take into account experimental imperfections by considering losses in the probe mode both before and after the sensor as well as losses in the reference mode.  Finally, the probe beam and reference beam (for the case of the bTMSS) are measured to perform the estimation. It is important to note that in performing comparisons between the transmission-based and phase-based schemes and the use of quantum and classical states, we keep the number of photons probing the sensor, $N$, fixed, as it represents the resource for the estimation.

We focus our attention on the use of two-mode squeezed states of light, as these quantum states have already been shown to reduce the uncertainty in the estimation of  phase~\cite{braunstein1994statistical,anderson2017optimal,anderson2017phase,Gong2017,Hu2018,You2019} and transmission~\cite{Invernizzi2011,woodworth2020transmission} below the SNL. Such quantum enhancements result from the presence of quantum correlations between both the amplitude quadratures and the phase quadratures of the two modes.  These correlations are due to the parametric process that is used to generate them~\cite{Loudon1987,Teich1989,Drummond2004}.
When not seeded, the parametric process grows from spontaneous emission and the average values of the field quadratures remain zero. It is possible to obtain larger mean values by displacing one or two of the modes before or after the parametric processor. In the limit in which the displacement is much greater than the quadrature fluctuations, the state becomes a bTMSS. The high intensity of such a bright state improves the absolute sensitivity of the sensor by probing with a large number of photons. Moreover, compared to other quantum states such as NOON states~\cite{Gilbert2008,Rubin2007}, the bTMSS are more robust to optical losses. Additionally, having two modes allows for differential measurements that can reduce or eliminate classical technical noise present in experimental implementations.

While it is well known that sensing schemes based on phase and transmission estimation can have different sensitivities~\cite{kabashin2009phase,lodewijks2012boosting,humer2012phase,yesilkoy2019optical,deng2017phase}, and comparisons between these schemes have been done for some resonance sensors~\cite{humer2012phase,yesilkoy2019optical,deng2017phase}, these previous studies have not been performed at their fundamental sensitivity limits. To perform a fair and absolute comparison between the transmission- and phase-based schemes, we calculate their fundamental sensitivity bound $S^{\textrm{b}}(\lambda|X)$, which is obtained when parameter $X$ is estimated at its QCRB. Thus, $S^{\textrm{b}}(\lambda|X)$ is inversely proportional to the fundamental minimum uncertainty in the estimation of the physical quantity using a given probing state of light. The QCRBs for transmission and phase estimation are independent of the measurement strategy and will be discussed in the next section. The fundamental sensitivity bounds can then be calculated by taking into account the corresponding transfer function, as will be discussed in section~\ref{comparison}.  Such an approach makes it possible to establish under which conditions a given scheme provides better sensitivities as well as the degree to which each scheme can take advantage of quantum resources.

\section{QCRB for Transmission and Phase Estimation}\label{QCRB}
The uncertainty in the estimation of a given parameter is fundamentally limited by the QCRB, which depends on the properties of the state of light used in the sensing configuration and the unitary transformation that is used to model the sensor. To calculate the QCRB we model the resonance sensor as a beam splitter~\cite{Kim2002} with intensity transmission $T$ followed by a phase rotation  $\phi$. For the resonance sensors that we consider here, both the transmission of the beam splitter and the phase rotation are wavelength dependant parameters as given by the transfer function of the sensor. Moreover, we model losses external to the sensor with beam splitters of constant transmissions  $\eta_{p1}$ and $\eta_{p2}$  for the probe mode before and after the sensor, respectively, and $\eta_r$ for the reference mode, as shown in figure~\ref{Fig:Del_nShift}(b).
We further assume that such losses do not introduce any additional phase rotations.

As we have previously shown~\cite{woodworth2020transmission}, the QCRB for transmission estimation with a bTMSS is given by
\be \label{EQ:QCRB_TMSS_Transmission}
\var{T}_Q^\textrm{bTMSS}\ge\frac{T}{\eta_{p2}N}-\frac{T^2}{N}\eta_{p1} D_r [1-\sech(2s)]
\ee
in the limit in which the vacuum terms can be neglected. In this equation, the number of probing photons $N$ is given by the product of the transmission before the sensor ($\eta_{p1}$) and the number of photons generated by the source, $s$ is the squeezing parameter that controls the quantum correlations between the quadratures~\cite{Loudon1987,Teich1989,Drummond2004}, and $D_r$, which contains the effect of losses in the reference arm, is defined as
\be \label{Conjugate_detection}
D_r=\frac{(2\eta_r-1)[1+2\sinh^2(s)]}{1+2\eta_r\sinh^2(s)}.
\ee
As can be seen from this result, an increase of quantum correlations in the bTMSS, characterized by $s$, leads to a reduction of the QCRB and thus a better sensitivity. In the extreme limit of perfect quantum correlations, $s\rightarrow\infty$, the QCRB for the bTMSS tends to the QCRB for the Fock state, known to have the lowest possible QCRB for transmission estimation of any quantum state~\cite{Adesso2009}. It is important to note that the QCRB for transmission always has a local minimum at a transmission of zero and for large enough values of $s$ at a transmission of one, as can be seen from equation~\eqref{EQ:QCRB_TMSS_Transmission}.

Similarly, as we show in \ref{App:QCRB}, the QCRB for phase estimation using a bTMSS in the presence of external losses is given by
\be \label{EQ:QCRB_TMSS_Phase}
\var{\phi}_Q^\textrm{bTMSS}\ge\frac{1}{4T\eta_{p2}N}-\frac{1}{4N}\eta_{p1} D_r [1-\sech(2s)],
\ee
given the assumption that there exists an external phase reference for each mode of the bTMSS~\cite{Jarzyna2012,You2019} and that the vacuum contribution is negligible.
This result is an extension of previous results~\cite{anderson2017optimal}, as it takes into account optical losses in the probe and reference beams.
As can be seen, external losses affect the QCRB for transmission, equation~\eqref{EQ:QCRB_TMSS_Transmission}, and phase, equation~\eqref{EQ:QCRB_TMSS_Phase}, in the same way. On the other hand, the two bounds have a different dependence on the transmission through the resonance sensor. For example, as opposed to the QCRB for transmission, the QCRB for phase provides maximum sensitivity only for transmissions of unity, which corresponds to the case when all the photons carrying information are measured.

The corresponding QCRBs for transmission and phase estimation with coherent states can be obtained from equations~\eqref{EQ:QCRB_TMSS_Transmission}~and~\eqref{EQ:QCRB_TMSS_Phase} by setting the squeezing parameter to zero, $s=0$. In this limit there is no parametric amplification and only the displaced vacuum states, which correspond to the coherent states, are left.
Although the reference mode may also be displaced, in the absence of the parametric process this second uncorrelated coherent state would only add noise to the estimation and as such the reference mode does not contribute when calculating the QCRB for the coherent state~\cite{woodworth2020transmission}.  This effectively simplifies to a configuration with a single mode coherent state probing the sensor.
The QCRBs for transmission and phase estimation with a coherent state are then given by
\bea
\var{T}_Q^\textrm{SNL}&\ge&\frac{T}{\eta_{p2}N},\label{EQ:QCRB_CS_T}\\
\var{\phi}_Q^\textrm{SNL}&\ge&\frac{1}{4T\eta_{p2}N}, \label{EQ:QCRB_CS_Phi}
\eea
respectively. Equations~\eqref{EQ:QCRB_CS_T} and~\eqref{EQ:QCRB_CS_Phi} are just the first terms on the right hand side of equations~\eqref{EQ:QCRB_TMSS_Transmission} and~\eqref{EQ:QCRB_TMSS_Phase}, which give the corresponding QCRBs when probing with bTMSS, respectively.
Thus, the second terms in equations~\eqref{EQ:QCRB_TMSS_Transmission} and~\eqref{EQ:QCRB_TMSS_Phase} are the ones that lead to the quantum advantage with bTMSS.

To characterize the quantum enhancement with respect to the SNL that can be obtained when using quantum states to probe the sensor, we define the quantum enhancement factor (QEF) as the ratio of the QCRB when probing with coherent states to the QCRB when probing with bTMSSs with equal resources.  Thus, the QEF for both transmission and phase estimation is given by
\be
\textrm{QEF}(X)=\frac{\Delta^2 X_{\textrm{QCRB,SNL}}}{\Delta^2 X_{\textrm{QCRB,bTMSS}}}=
\left\{1-T\eta_{p1}\eta_{p2}D_r[1-\sech(2s)]\right\}^{-1},
\label{Eq:QEF}
\ee
where a value greater than one corresponds to lower uncertainties in the estimation of transmission or phase when probing with a bTMSS as compared to a coherent state. Given that the QEFs for both transmission and phase estimation are the same for a given transmission $T$, both cases are able to take equal advantage of quantum resources. It is important to note that a quantum enhancement only happens if losses in the reference beam are small enough such that $\eta_r>1/2$.  For the case in which $\eta_r<1/2$, we have that $D_r<0$, which makes the QEF drop below one.  In this case there is no advantage in using bTMSSs over  coherent states, as will be further discussed in section~\ref{Losses}.

\section{Quantum-Enhanced Sensitivity of Resonance Sensors}\label{comparison}
In order to fully  characterize the sensitivity $S(\lambda|X)$ of the resonance sensor, we also need to take into account its transmission or phase spectrum. In general, the spectrum is characterized by the sensor's transfer function $t(\lambda)$, which defines the complex response of the sensor as a function of wavelength. We consider a general transfer function whose intensity transmission transfer function can be written as:
\be
\textrm{T}(\lambda) = |t(\lambda)|^2= T_{\textrm{off}} + (T_{\textrm{res}} - T_{\textrm{off}}) T_0(\lambda),
\label{TRF}
\ee
where $T_{\textrm{off}}$ and $T_{\textrm{res}}$ are the far-off-resonance and on-resonance transmissions, respectively, and $T_0(\lambda)$ defines the intensity resonance lineshape.
The parameters $T_{\textrm{res}}$ and $T_{\textrm{off}}$ in equation~\eqref{TRF} make it possible to define a transfer function characterized by a peak or dip resonance responses when $T_{\textrm{res}}>T_{\textrm{off}}$ or $T_{\textrm{res}}<T_{\textrm{off}}$, respectively. To fully define the transfer function, we consider a resonance lineshape, $T_0(\lambda)$, that has unit transmission on resonance and goes to zero away from the resonance. Due to its broad applicability, we focus on a Lorentzian lineshape of the form
\be \label{Lorentz}
  T_0(\lambda)=\left|\frac{1}{\frac{\lambda-\lambda_0}{\Delta L}+i}\right|^2,
\ee
where $\lambda_0$ is the resonance wavelength, $\Delta L$ is the half-width-at-half-maximum (HWHM), and $i=\sqrt{-1}$.

Since we only consider sensors with a linear response, their transmission and phase responses are related through the Kramers-Kronig relations~\cite{kop1997kramers,gralak2015phase}. As we show in~\ref{App:Transfer}, for arbitrary values of $T_{\textrm{res}}$ and $T_{\textrm{off}}$, the transmission and phase transfer functions for a Lorentzian lineshape are given by
\bea
T(\Lambda)&=&T_\textrm{off}+\frac{T_\textrm{res}-T_\textrm{off}}{1+\Lambda^2},\label{EQ:TLorentz}\\
\phi(\Lambda)&=&\arctan\left[\frac{\Lambda^2\sqrt{T_\textrm{off}}+\sqrt{T_\textrm{res}}}{\Lambda\left(\sqrt{T_\textrm{res}}-\sqrt{T_\textrm{off}}\right)}\right],\label{EQ:PhaseLorentz}
\eea
where $\Lambda=(\lambda-\lambda_0)/\Delta L$ is the generalized wavelength, defined to make the transfer function independent of the HWHM and the resonance wavelength of the Lorentzian lineshape.
With this definition, the sensitivity of the resonance sensor will be re-scaled as $S(\lambda|X)=(\Delta L)^{-2}S(\Lambda|X)$, which represents another common factor in our comparison between the sensing schemes based on transmission and phase.

We can now combine the QCRB for transmission or  phase estimation given by  equations~\eqref{EQ:QCRB_TMSS_Transmission} and~\eqref{EQ:QCRB_TMSS_Phase}, respectively, with the corresponding derivative of the transfer function to show that the fundamental sensitivity limit for a resonance sensor with a Lorentzian transmission lineshape takes the form
\bea
 S^\textrm{b}(\Lambda|T)&=&N \Bigg\{\frac{(1 + \Lambda^2)^3 (\Lambda^2T_{\textrm{off}} + T_{\textrm{res}})}{4\Lambda^2\eta_{p2}(T_{\textrm{off}}-T_{\textrm{res}})^2}-\frac{(1 + \Lambda^2)^2 (\Lambda^2 T_{\textrm{off}}+T_{\textrm{res}})^2}{4\Lambda^2 (T_{\textrm{off}}-T_{\textrm{res}})^2}\eta_{p1}D_r[1-\sech(2s)]\Bigg\}^{-1},\label{Eq:sensitivityT}\\
 S^\textrm{b}(\Lambda|\phi)&=&N\Bigg\{\frac{\left(1+\Lambda^2\right)^3\left(\Lambda^2T_{\textrm{off}}+T_{\textrm{res}}\right)}{4\eta_{p2}\left(\sqrt{T_{\textrm{off}}}-\sqrt{T_{\textrm{res}}}\right)^2\left(\Lambda^2\sqrt{T_{\textrm{off}}}-\sqrt{T_{\textrm{res}}}\right)^2}
 \nonumber\\&&-\frac{\left(1+\Lambda^2\right)^2\left(\Lambda^2T_{\textrm{off}}+T_{\textrm{res}}\right)^2}{4\left(\sqrt{T_{\textrm{off}}}-\sqrt{T_{\textrm{res}}}\right)^2\left(\Lambda^2\sqrt{T_{\textrm{off}}}-\sqrt{T_{\textrm{res}}}\right)^2}\eta_{p1}D_r[1-\sech(2s)]\Bigg\}^{-1},\label{Eq:sensitivityPhi}
\eea
for the transmission- or phase-based scheme, respectively, when the sensor is probed with a bTMSS.
Due to the scaling of these sensitivities with the number of probing photons, $N$, we can define the sensitivity per probing photon $S^\textrm{b}(\Lambda|X)/N$ to make the analysis and comparison  independent of the number of probing photons.

\begin{figure}
    \centering
    \includegraphics{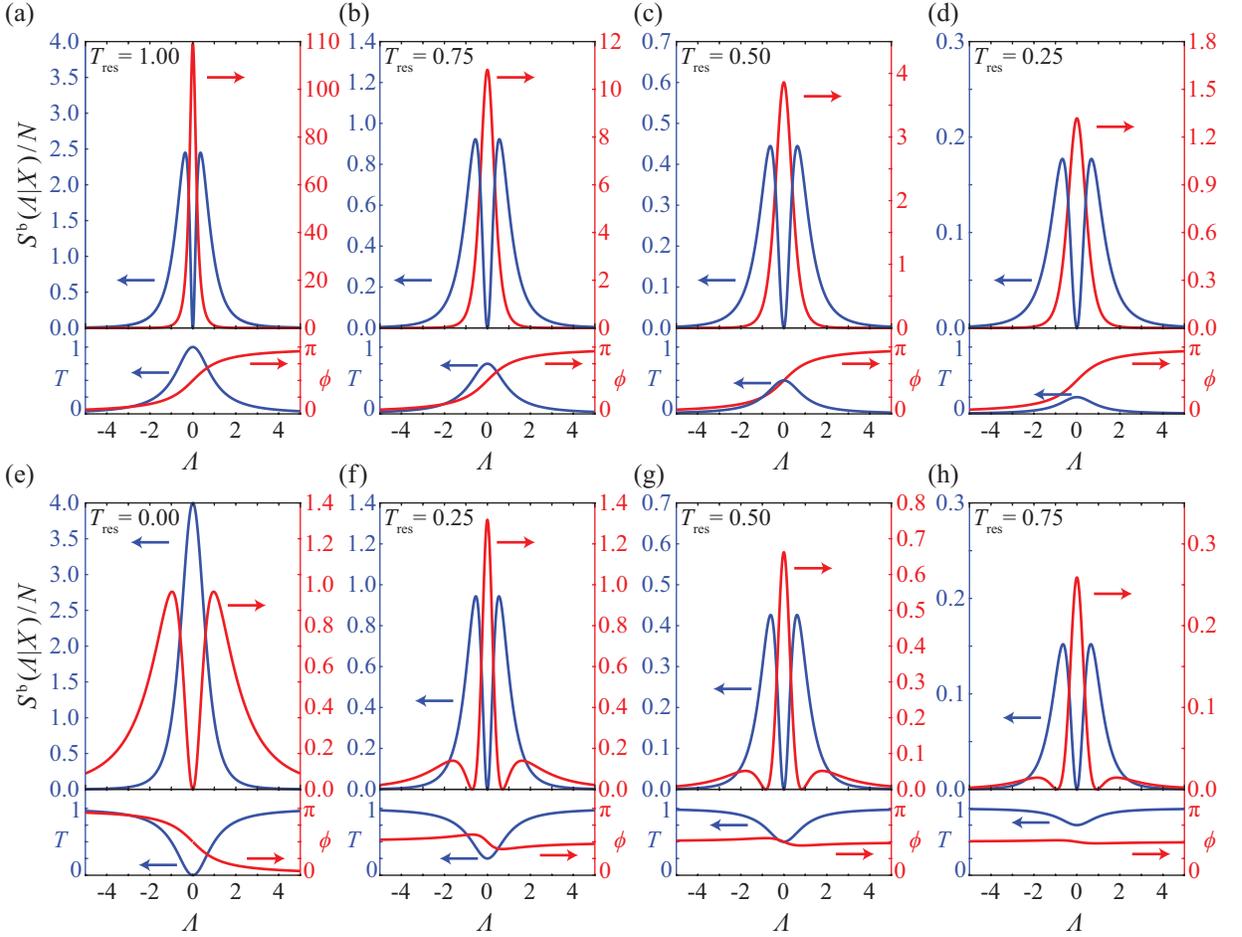}
    \caption{Sensitivity per probing photon ($S^{b}(\Lambda|X)/N$) for resonance sensors with a transmission-based (blue) and phase-based (red) schemes.  We assume a Lorentzian lineshape for the resonance, as shown in the lower portion of each figure, and consider different values of $T_\textrm{res}$ and $T_\textrm{off}$ to study the behavior of a transfer function with either a peak, (a) through (d), or a dip, (e) through (h), resonance. The maxima in these traces occur at the wavelengths at which the sensitivity is maximized. The exact wavelength at which this happens results form an interplay of between the QCRB and the slope of the corresponding transfer function.}
    \label{fig:EQFI_Lorentzian}
\end{figure}

In order to compare the two sensing schemes and the effects of having a resonance response with a peak or a dip, we plot the sensitivity per probing photon, $S^{\textrm{b}}(\Lambda|X)/N$, as a function of  wavelength for different values of $T_{\textrm{res}}$ and $T_{\textrm{off}}$ in figure~\ref{fig:EQFI_Lorentzian}. The bottom portion of each figure shows the corresponding transmission and phase transfer functions. For this comparison, we specialize to the case of no losses external to the sensor, that is $\eta_{p1}=\eta_{p2}=\eta_r=1$. We will address the effect of losses in section~\ref{Losses}. Furthermore, we consider a bTMSS with a squeezing parameter of $s=2$, which corresponds to $\sim -14.5$~dB of intensity-difference squeezing.
Such levels of squeezing have been experimentally generated for single-mode squeezed states~\cite{Vahlbruch2016}, and are within reach for the case of bTMSSs.

As can be seen from the blue traces in figures~\ref{fig:EQFI_Lorentzian}(a)-(d), for the transmission-based scheme for  a transfer function with a peak response, the maximum sensitivity is not achieved at the origin.  This results from having the derivative of the transmission transfer function be zero on resonance. Instead, the wavelength at which the maximum sensitivity occurs results from an interplay between the QCRB and the transmission slope of the transfer function. For example, for the case of a perfect peak shown in figure~\ref{fig:EQFI_Lorentzian}(a), the local minima of the QCRB for transmission occur at $T=0$ or $T=1$, where the slopes of the transmission transfer function are zero, while the steepest slopes of the transmission transfer function occur at $|\Lambda|=\pm1/\sqrt{3}$.

As expected, as the maximum resonance transmission through the sensor decreases, $T_{\textrm{res}}\rightarrow0$, the sensitivity also decreases. For the case in which the transmission transfer function has a dip resonance with a Lorentzian lineshape, figures~\ref{fig:EQFI_Lorentzian}(e)-(h), the overall behavior is similar to the case of a peak transmission except for a perfect dip, $T_{\textrm{res}}=0$.  As can be seen in figure~\ref{fig:EQFI_Lorentzian}(e), for the perfect dip the maximum sensitivity is achieved on exact resonance, $\Lambda=0$, given that for this case equation~\ref{EQ:QCRB_TMSS_Transmission} shows that there is no uncertainty in the estimation of transmission. This is no longer the case, however, when $T_{\textrm{res}}$ deviates from zero. When this happens the QCRB deviates from zero faster than the increase in slope of the transmission transfer function around resonance, which leads to a splitting of the maximum of the sensitivity per probing photon to two peaks that occurs at wavelength that are shifted symmetrically with respect to the resonance wavelength,
as can be seen from the blue traces in~figures~\ref{fig:EQFI_Lorentzian}(f)-(h).

In the limit of infinite squeezing, $s\rightarrow\infty$, the QCRB for transmission when probing with a bTMSS becomes symmetric about $T=0.5$,  as we can see from equation~\eqref{EQ:QCRB_TMSS_Transmission}. This means that the uncertainty in the estimation of transmission for $T=0.5-x$ and $T=0.5+x$, where $0\le x \le0.5$, have the same QCRB bound.
This leads to a symmetry in the sensitivity for the transmission-based scheme such that complementary  transmission values, $T_{\textrm{res}}\rightarrow1-T_{\textrm{res}}$ and $T_{\textrm{off}}\rightarrow1-T_{\textrm{off}}$, have the same sensitivity.
If we had considered the case of infinite squeezing for figure~\ref{fig:EQFI_Lorentzian}, then each column would have had the same sensitivity per probing photon as a function of wavelength for the transmission-based scheme (blue traces). This behavior can already be seen for $s=2$ in the last 3 columns of figure~\ref{fig:EQFI_Lorentzian}, where the sensitivity for these complementary pairs is nearly the same. However, since infinite squeezing is not possible, this symmetry in complementary transmission values is lifted and results in better sensitivity values for the case with lower transmission at the wavelength of maximum sensitivity.  That is, the last two columns for the peak resonance, figures~\ref{fig:EQFI_Lorentzian}(c)-(d), have higher maximum sensitivity than the corresponding ones for the dip resonance, figures~\ref{fig:EQFI_Lorentzian}(g)-(h), as they have lower transmission at the wavelengths at which the blue curve is maximized.  On the other hand, for the first column, the dip resonance, figure~\ref{fig:EQFI_Lorentzian}(e), has a better sensitivity than the peak resonance, figure~\ref{fig:EQFI_Lorentzian}(a). As a result, when no power is lost to the resonance sensor such that two outputs are accessible, one with a peak and the other with a dip in the transmission spectrum as is the case with an optical cavity for instance, the output with a lower transmission at the optimal operational wavelength will provide a better absolute sensitivity for the transmission-based scheme.
However, as we can see from equation~\eqref{Eq:QEF}, while such a lower transmission results in better absolute sensitivity it also leads to a lower level of quantum enhancement.  Thus, in the limit of small $T_\textrm{res}$, there will not be a significant quantum enhancement with the use of a bTMSS over a coherent state.

Next, we consider the sensitivity per probing photon as a function of wavelength for the phase-based scheme, see red traces in figure~\ref{fig:EQFI_Lorentzian}. As opposed to the transmission-based scheme,  for the phase-base scheme the steepest slopes for the phase transfer function for a Lorentzian lineshape occur at the resonance wavelength, $\Lambda=0$. For the case of a peak resonance, figures~\ref{fig:EQFI_Lorentzian}(a)-(d), the transmission is maximum on resonance and thus the QCRB for phase estimation has its lowest value, which leads to the maximum sensitivity occurring at the resonance wavelength. As $T_{\textrm{res}}$ decreases, the sensitivity of the phase-based scheme with a peak resonance is also reduced as the QCRB for phase estimation increases, but the optimum wavelength remains at the resonance wavelength.
For the case of a dip resonance, the maximum sensitivity is still at the resonance wavelength, as long as $T_{\textrm{res}}\ne0$, due to the sharp slope of the phase transfer function dominating over a larger QCRB for phase estimation, as can be seen from the red traces in figure~\ref{fig:EQFI_Lorentzian}(f)-(h). It is important to note that for these cases the phase transfer function exhibits two inflection points that lead to generalized wavelengths with zero sensitivity and two side lobes around the resonance frequency
that occur at wavelengths further away from resonance than the inflection points. Furthermore, as the $T_\textrm{res}$ decreases, the phase difference between the maximum and minimum values of the phase transfer function increases while their wavelength separation decreases. This leads to an increased slope around the resonance that leads to an increase in height and narrowing of the peak in the sensitivity at the resonance wavelength.  Additionally, this behavior leads to a better sensitivity on resonance than at the wavelengths of the side lobes where the transmission is higher.

For $T_\textrm{res}=0$ the phase difference between the inflection points maximizes while the wavelength separation minimizes and the two points become one as the phase winds back on itself.
While this leads to an increased slope on resonance, the transmission is zero at this point, which makes the QCRB go to infinity.  As a result, the sensitivity per probing photon goes to zero and the best sensitivity occurs at the wavelengths of the side lobes, as shown in figure~\ref{fig:EQFI_Lorentzian}(e).
For the phase-based scheme in the limit of infinite squeezing, $s\rightarrow \infty$, given that the uncertainty in the estimation of phase increases monotonically as $T$ decreases there is no symmetry for complementary transmissions as in the case of the transmission-based scheme.
Therefore, for sensors where two outputs, one with a peak resonance and the other with a dip resonance, are accessible, the output with the peak resonance will always exhibit the best sensitivity for the phase-based scheme.

\begin{figure}
    \centering
    \includegraphics{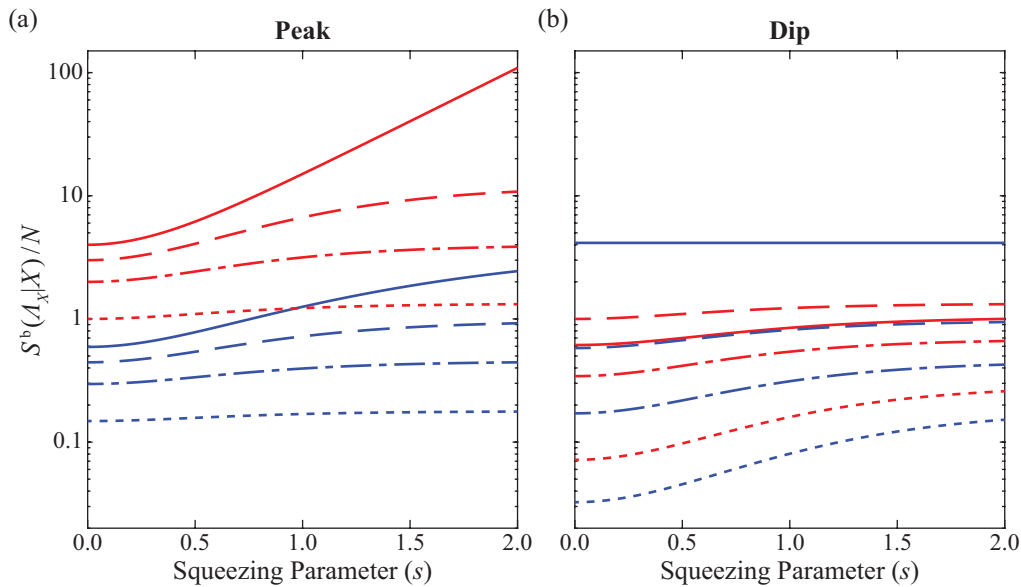}
    \caption{Maximum sensitivity per probing photon for transmission (blue) and phase (red) based schemes for resonance sensors with a Lorentzian lineshape, for peak resonances with $T_{\textrm{off}}=0$ (left), and dip resonances with $T_{\textrm{off}}=1$ (right). The $|T_{\textrm{off}}-T_{\textrm{res}}|$ equals $1.00$ for solid, $0.75$ for dashed, $0.5$ for dot-dashed, and $0.25$ for dotted lines, which correspond to the same transfer functions considered in figure~\ref{fig:EQFI_Lorentzian}. }
    \label{fig:MAX_EQFI_vs_s}
\end{figure}

Although the quantum enhancements in the estimation of transmission and phase are the same, as can be seen from equation~\eqref{Eq:QEF}, the quantum enhancement in sensitivity for the resonance sensor is different for each scheme.  This is a result of the difference in the wavelength dependence of the transmission and phase transfer functions. This leads to an optimum wavelength at which the maximum sensitivity occurs to be different for the transmission-based and phase-based schemes, as can be seen in figure~\ref{fig:EQFI_Lorentzian}. In order to perform a fair comparison between the two sensing schemes, we compare the maximum values of $S^{b}(\Lambda|X)/N$ irrespective of the wavelength at which they occur. That is, we assume that the resonance sensors is always being probed at its optimal wavelength, which changes with the sensing scheme and the squeezing parameter $s$ of the bTMSS used to probe the sensor.
Figure~\ref{fig:MAX_EQFI_vs_s} shows the maximum values of the sensitivity per probing photon for each sensing scheme as a function of squeezing parameter.  As shown in figure~\ref{fig:MAX_EQFI_vs_s}(a), for resonance sensors with a peak resonance and a Lorentzian lineshape, the sensitivity for both sensing schemes increases  with increasing level of squeezing parameter. However, for the same $|T_{\textrm{off}}-T_{\textrm{res}}|$, the phase-based scheme (red traces) always provide better sensitivities than the transmission scheme (blue traces), even for coherent states, $s=0$.

The behavior is similar for the case of a dip resonance except when $T_\textrm{res}=0$, solid lines in figure~\ref{fig:MAX_EQFI_vs_s}(b). For this case
the transmission-based scheme has a better sensitivity than the phase-based one, consistent with the behavior shown in figure~\ref{fig:EQFI_Lorentzian}.  However, the transmission-based scheme does not show any enhancements over a coherent state when using a bTMSS even with higher squeezing parameters, as can be seen from the flat solid blue trace in figure~\ref{fig:MAX_EQFI_vs_s}(b), given that the maximum sensitivity occurs on resonance and that the QEF for the estimation of transmission goes to one for a transmission of zero.

\begin{figure}
    \centering
    \includegraphics{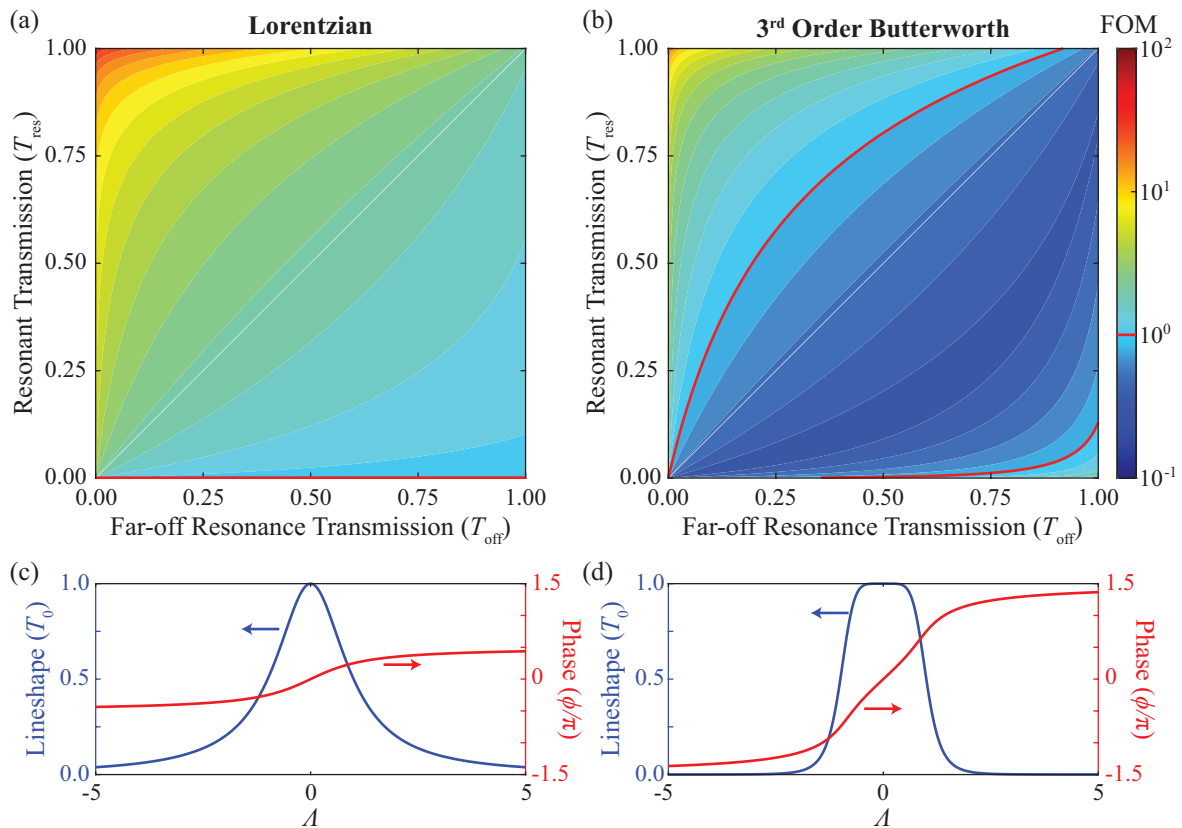}
    \caption{Figure of merit (FOM) as a function of $T_{\textrm{res}}$ and $T_{\textrm{off}}$ for (a) a Lorentzian resonance lineshape and (b) a Butterworth resonance lineshape with $m=3$ assuming lossless conditions and $s=2$. For the Lorentzian lineshape, the FOM is greater than one for most values of $T_{\textrm{res}}$ and $T_{\textrm{off}}$; while for the Butterworth lineshape, sharper transmission slopes allows the transmission-based schemes to outperform the phase-based schemes for a larger parameter space. The red lines indicate $\textrm{FOM}=1$, which correspond to the transition from one scheme dominating over the other. Transfer functions for perfect peak resonances for (c) Lorentzian and (d) Butterworth ($m=3)$ lineshapes. Blue (red) traces correspond to the transmission (phase) spectrum.}
    \label{fig:FOM_Butterworth}
\end{figure}

To quantify the comparison between the sensitivities of the  phase- and transmission-based schemes, we define the figure of merit (FOM) as the ratio between the sensitivities at the optimal wavelength of the two schemes, that is
\be\label{SensitivityRatio}
  \textrm{FOM}=\frac{\underset{\Lambda}{\textrm{max}}\:S^\textrm{b}(\Lambda|\phi)}{\underset{\Lambda}{\textrm{max}}\:S^\textrm{b}(\Lambda|T)}.
\ee
A FOM greater than one indicates that the phase-based scheme outperforms the transmission-based one.
Figure~\ref{fig:FOM_Butterworth}(a) shows the FOM for different values of $T_\textrm{res}$ and $T_\textrm{off}$ for a Lorentzian lineshape and a squeezing parameter of $s=2$.
For almost all values of $T_{\textrm{res}}$ and $T_{\textrm{off}}$, the phase-based scheme is more sensitive than the transmission-based one. For resonance sensors with a Lorentzian lineshape, the FOM becomes less than or equal to one only in the limit of $T_{\textrm{res}}= 0$, as shown by the red line in figure~\ref{fig:FOM_Butterworth}(a).

The sensitivity of the resonance sensor, and consequently the FOM, depends on the lineshape of the transfer function. A Gaussian lineshape, for example, will have a FOM similar to the one for a Lorentzian lineshape due to a similar behavior of the transmission and phase transfer functions.
To illustrate the effect of the lineshape on the sensitivity of a resonance sensor for both schemes and the FOM we consider a Butterworth lineshape, which is a generalization of a Lorentzian lineshape that exhibits larger slopes in the transmission transfer function as the order increases. The transmission transfer function for the Butterworth lineshape takes the form
\be
T(\Lambda;m)=\frac{T_{\textrm{res}}-T_{\textrm{off}}}{1+\Lambda^{2m}}+T_{\textrm{off}},
\ee
where $m$ is the order of the Butterworth lineshape, while the phase transfer function, given by the red trace in  figure~\ref{fig:FOM_Butterworth}(d), is calculated numerically as shown in~\ref{App:Transfer}. Note that the first order, $m=1$, Butterworth lineshape simplifies to a Lorentzian one. As the order increases, the transmission lineshape tends towards a square one with the transmission flattening around the resonance, $\Lambda=0$, and steeper slopes around $|\Lambda|=1$ than those of a Lorentzian lineshape, as shown in figure~\ref{fig:FOM_Butterworth}(d) for $m=3$.

Figure~\ref{fig:FOM_Butterworth}(b) shows the FOM for all possible values of $T_{\textrm{res}}$ and $T_{\textrm{off}}$ for a third order Butterworth lineshape, $m=3$, with $s=2$.  Note that the transition between a peak and dip in the transmission transfer function occurs along the diagonal $T_{\textrm{res}}=T_{\textrm{off}}$, with a peak resonance corresponding to the upper left region and a dip resonance corresponding to the lower right region. For the third order Butterworth lineshape, the FOM falls below one for a significant range of $T_{\textrm{res}}$ and $T_{\textrm{off}}$ values around the transitions between the peak and dip resonance, outlined by the red lines in figure~\ref{fig:FOM_Butterworth}(b).  This region corresponds to the parameter space for which the transmission-based scheme outperforms the phase-based one. This is a much larger region than the one for a Lorentzian lineshape and illustrates the interplay between the QCRB for transmission or phase and the corresponding slope of the transfer function.  Thus, for an arbitrary resonance sensor, such as one with an asymmetric transmission response, it cannot be assumed that the phase-based scheme will always result in a better sensitivity than the transmission-base scheme.

\section{Effect of Losses on Sensitivity}\label{Losses}

To study more realistic operational conditions, we now consider the effect of optical losses external to the resonance sensor on the sensitivity. We take into account sources of optical loss, such as imperfect optical elements and photo-detectors, by modeling them with a beam splitter with intensity transmissions $\eta_{p1}$ and $\eta_{p2}$ before and after the resonance sensor, respectively, in the path of the probing beam and with intensity transmission $\eta_{r}$ in the path of the reference beam, as outlined in figure~\ref{Fig:Del_nShift}(b). As can be seen from equations~\eqref{EQ:QCRB_TMSS_Transmission}~and~\eqref{EQ:QCRB_TMSS_Phase}, these losses lead to an increase in the QCRB for the estimation of transmission and phase.
Thus, as expected, losses external to the resonance sensor always lead to a reduction in its sensitivity for both sensing schemes.

In studying the effect of losses, we need to consider that different states of light are affected differently by optical losses and that the impact of losses on the sensitivity depends on where they happen. For example, the photon statistics of a coherent state remain unchanged after optical losses with the only impact being a reduction in the amplitude of the coherent state. Thus, when probing the resonance sensor with a coherent state, losses before the sensor will have no impact on the sensitivity, as the number of photons probing the sensor represents the resource for the estimation and is kept constant. Moreover, since there is no reference beam when probing with a coherent state, the sensitivities will be independent of $\eta_r$. However, losses after the sensor will have an impact on the sensitivity as photons containing information from the interaction with the sensor will be lost. On the other hand, when probing with bTMSS all three sources of loss will lead to a reduction in sensitivity.

In comparing the sensing schemes based on transmission or phase, we focus on the study of the effect of losses on the quantum enhancement that can be achieved with either scheme rather than their effect on the absolute sensitivities, as there is no advantage in using a quantum state if it does not lead to a quantum enhancement. To do so, we define the effective quantum enhancement factor (EQEF) as the ratio of the sensitivities at their optimal wavelengths for the sensing scheme of interest when probing with bTMSS to the one when probing with a coherent state, which provides the SNL, while keeping the number of photons probing the sensors constant, that is
\be
\textrm{EQEF}(X)=\frac{\underset{\Lambda}{\max}\:S^{\textrm{b}}(\Lambda|X)_{\textrm{bTMSS}}}{\underset{\Lambda}{\max}\:S^{\textrm{b}}(\Lambda|X)_{\textrm{SNL}}}.
\label{Eq:EQEF}
\ee
It is important to note that the wavelength at which the maximum sensitivity is achieved will be different for the bTMSS and the coherent state. Compared to the quantum enhancement factor defined in equation~\eqref{Eq:QEF}, the EQEF includes not only the QCRB of the estimation parameter $X$, but also the response of the sensor.

\begin{figure}
    \centering
    \includegraphics{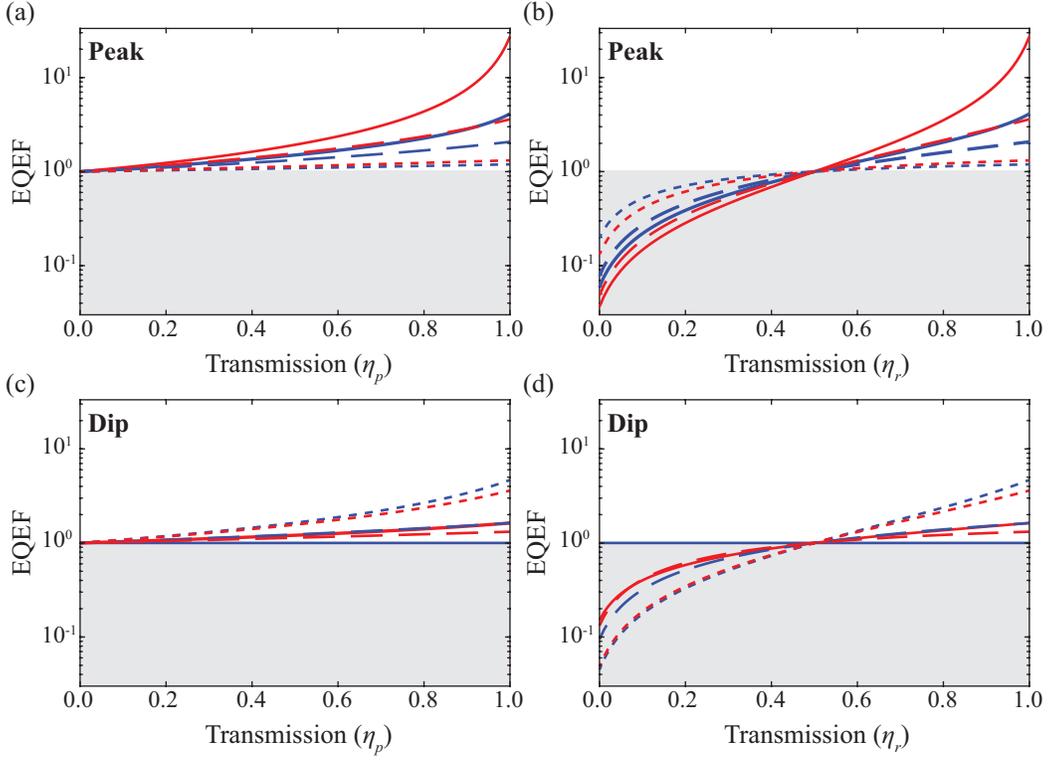}
    \caption{The effective quantum enhancement factor (EQEF) for transmission (blue) and phase (red) based schemes for a Lorentzian lineshape resonance sensor, for peak resonances, (a) and (b), with $T_{\textrm{off}}=0$, and dip resonances, (c) and (d), with $T_{\textrm{off}}=1$. The $|T_{\textrm{off}}-T_{\textrm{res}}|$ equals $1.00$ for solid, $0.75$ for dashed, and $0.25$ for a dotted line, which correspond to the transfer functions shown in figure~\ref{fig:EQFI_Lorentzian} expect for the $0.50$ difference that was omitted for figure clarity. The left hand plots (a) and (c) show the reduction of the EQEF as a function of total probe losses external to the resonance sensor, with $\eta_{p}=\eta_{p1}\eta_{p2}$.  The right hand plots (b) and (d) show the reduction in EQEF due to losses in the reference beam.}
    \label{fig:EQEF_loss}
\end{figure}

The behavior of the EQEF in the presence of optical losses is shown in figure~\ref{fig:EQEF_loss} for different parameters of a peak or dip Lorentzian lineshape for a bTMSS with $s=2$. In general, losses in the probe beam, whether before or after the sensor, lead to a change in the photon statistics of the probe beam. Losses after the sensor also lead to a loss of information encoded in the light as a result of the interaction with the sensor. However, probe losses before and after the sensor, $\eta_{p1}$ and $\eta_{p2}$, have the same mathematical effect on the EQEF, as can be shown from equations~\eqref{Eq:sensitivityT} and~\eqref{Eq:sensitivityPhi}. As expected, as losses external to the sensor increase the degree of quantum enhancement is reduced and the EQEF tends monotonically towards unity for both the transmission- and phase-based sensing schemes, as shown in figures~\ref{fig:EQEF_loss}(a) and (c). For the particular case of resonance sensors with low transmissions, such as a dip resonance when $T_{\textrm{res}}\rightarrow0$, the transmission QCRB for the coherent state and the bTMSS tend to the same value independent of external losses.  As a result, there is no significant advantage to using quantum states over coherent states and the EQEF for the transmission-based scheme, blue solid line in figure~\ref{fig:EQEF_loss}(c), is fairly insensitive to losses in the probe arm.  The same behavior is not seen in this case for the phase-based scheme, red solid line in figure~\ref{fig:EQEF_loss}(c), for which there is a quantum advantage that degrades as  losses in the probe arm increase.

While for our sensing configuration the reference beam does not carry any information about the parameter of interest, losses in the reference beam lead to a reduction of the quantum correlations in the bTMSS. As a result, such losses lead to an increase in the uncertainty in the estimation of transmission or phase and thus a reduction in sensitivity. When the losses in the reference beam exceeds 50\%, $\eta_r<1/2$, the uncertainty in the estimation of transmission and phase with a bTMSS exceeds the corresponding ones with a coherent state, as shown in equation~\eqref{Eq:QEF}. This results from the fact  that each mode of the bTMSS by itself has more noise than a coherent state. Thus, the EQEF drops below one, as shown in figures~\ref{fig:EQEF_loss}(c) and (d), which means that probing with a coherent state would lead to a higher sensitivity than probing with a bTMSS. The only exception to this behavior is for the transmission-based scheme in the case of a full-dip Lorentzian resonance ($T_\textrm{res}=0$ and $T_{\textrm{off}}=1$), as the EQEF without losses is already at the level of a coherent state and the bound becomes insensitive to losses in the reference beam, see solid blue trace in figure~\ref{fig:EQEF_loss}(d).

It is important to note that while the phase-based scheme always has higher sensitivity than the transmission based scheme, except when $T_\textrm{res}=0$, for a Lorentzian lineshape (see figure~\ref{fig:EQFI_Lorentzian}), it does not always lead to the largest quantum enhancement. For example, we can see in figures~\ref{fig:EQEF_loss}(c) and (d) that for $T_\textrm{off}=1.00$ and $T_\textrm{res}=0.75$, dashed lines, the transmission EQEF is larger than the phase EQEF.

\section{Measurement Strategies that Saturate the QCRB}\label{SensingSchemes}

While the theoretical fundamental bounds for the sensitivities of resonance sensors can be determined with the QCRB, it is still necessary to identify measurement techniques that can saturate those bounds to fully take advantage of the bTMSS. It has been shown that the QCRB for transmission and phase estimation can be reached with current detection techniques based on intensity or homodyne measurements~\cite{anderson2017optimal,woodworth2020transmission}.

For phase estimation, a phase sensitive detection method, such as an interferometer or homodyne detection (HD), is needed. In these techniques, changes in the phase of the beam under study are measured with respect to the phase of an external beam called the local oscillator (LO). For the case of a bTMSS, given that quantum noise reduction is observed in the phase-sum quadrature, a common approach to perform phase measurements is to use a HD for each of the modes. It has been shown that the bTMSS QCRB for phase estimation can be saturated with such an approach if one allows for gain in the electronic signal from one of the HDs before obtaining the difference~\cite{anderson2017optimal,gupta2018optimized}. The electronic gain is set to minimize the noise in the measured signal. Moreover, if the phase of the LOs for the probe and the reference arms are set such that the amplitude quadratures of the beams are measured, such  optimized HD can also be used to saturate the QCRB for transmission estimation.  As we show in~\ref{App:HD}, the optimized HD can saturate the QCRB for phase and transmission estimation even in the presence of losses external to the resonance sensor. Alternatively, for transmission estimation, the bTMSS QCRB can be saturated with transmission measurements by implementing an optimized intensity-difference
measurement~\cite{woodworth2020transmission}. In this approach two photodiodes are used to measure the intensity of the probe and reference beams.  As with the optimized HD, electronic gain is then used on the electronic signal from one of the photodiodes to minimize the measured noise in the difference signal.

\section{Conclusion}\label{conclusion}
We present a study of the fundamental sensitivity limits for estimating a quantity of interest (such as force, temperature, pressure, index of refraction, etc.) with optical resonance sensors based on the estimation of the change in transmission or phase of the light used to probe them. We obtain the fundamental limits by calculating the QCRB for each of the sensing approaches and show that an interplay between the QCRB and the transfer function of the resonance sensor determines the fundamental sensitivity that can be achieved with either sensing scheme. We show that for the case of a resonance sensor with a Lorentzian lineshape, the phase-based scheme outperforms the transmission-based scheme for almost all operational parameters.  However, due to the interplay between the QCRB and the transfer function, this is no longer the case when we consider lineshapes with steeper slopes in the transmission transfer function, such as a third order Butterworth lineshape. Furthermore, our results show that in the ideal case of a lossless configuration for a resonance sensor with a peak Lorentzian lineshape, the phase-based sensing scheme can provide over an order of magnitude higher sensitivity than the transmission-based scheme when proved with a bTMSS with $s=2$.

We also study the effect of losses on the quantum enhancement that can be achieved with each of the schemes and find, as expected, that optical losses degrade the level of quantum enhancement. There are two interesting things  worth noting.  First, the losses on the reference arm can have a significant impact on the quantum enhancement and in fact lead to sensitivities lower than those of a coherent state for losses larger than $50\%$. Second, while the phase-based scheme has a  higher sensitivity than the transmission-based scheme for a Lorentzian lineshape for most operational parameters, it does not always lead to the largest degree of quantum enhancement. Finally, we present measurement techniques that have been perviously shown to be able to saturate the bTMSS QCRB for transmission and phase estimation and thus offer experimentally accessible measurements to reach the ultimate sensitivity limits for resonance sensors.

\section{Acknowledgments}
This work was supported by the W. M. Keck Foundation and by the National Science Foundation (NSF) (Grant No. PHYS-1752938).

\appendix
\section{Lossy QCRB for Phase Estimation with a bTMSS}\label{App:QCRB}
To calculate the QCRB for phase estimation with a bTMSS in the presence of loss, we followed the technique used by \v{S}afr\'anek \textit{et al.} in~\cite{Safranek2015} for Gaussian states.  These states, such as the bTMSS, are fully characterized by the displacement vector, $\vec d$, and covariance matrix, $\boldsymbol\sigma$, which are defined as
\bea
\vec d&=&\big\langle\vec{\hat{A}}\big\rangle\\
\sigma_{i,j}&=&\big\langle\hat{A}^{}_i\hat{A}^\dagger_j+\hat{A}^\dagger_j\hat{A}^{}_i\big\rangle-2\big\langle\hat{A}_i\big\rangle\big\langle\hat{A}^\dagger_j\big\rangle,
\eea
where $\vec{\hat{A}}=\left(\hat a_p^{},\hat a_r^{},\hat a_p^\dagger,\hat a_r^\dagger\right)^T$. In the bright limit in which the displacement term dominates, the QCRB is given by
\be
\var{\phi}\ge\left(2\dot{\vec{d}}^{\: \dagger}\boldsymbol\sigma^{-1}\dot{\vec{d}}\:\right)^{-1},\label{eqn:QCRBderive}
\ee
where $\dot{\vec{d}}$ is the element-wise derivative of the displacement vector with respect to the phase.

To take into account losses in the calculation of the QCRB for phase estimation with a bTMSS, the following operator transformations are needed
\bea
\hat S^\dagger_{p,r}\hat a_p^{}\hat S^{}_{p,r}&=&\hat a_p^{}\cosh(s)-\hat a^\dagger_re^{i\theta}\sinh(s),\label{eqn:Transf1}\\
\hat S^\dagger_{p,r}\hat a_r^{}\hat S^{}_{p,r}&=&\hat a_r^{}\cosh(s)-\hat a^\dagger_pe^{i\theta}\sinh(s),\\
\hat B^\dagger_{i}(T_i)\hat a_{i}^{}\hat B^{}_{i}(T_i)&=&\sqrt{T_i}\hat a^{}_{i}+\sqrt{1-T_i}\hat a_\nu^{},\\
\hat D^\dagger_{p}(\alpha)\hat a_p^{}\hat D^{}_p(\alpha)&=&\hat a^{}_p+\alpha,\\
\hat D^\dagger_{r}(\beta)\hat a_r^{}\hat D^{}_r(\beta)&=&\hat a^{}_r+\beta,\\
\hat\Phi^\dagger_p(\phi)\hat a_p^{}\hat\Phi^{}_p &=& e^{i\phi}\hat a_p^{},\label{eqn:Transf6}
\eea
where $\hat S_{p,r}$ is the two-mode squeezing operator with squeezing parameter $s$ and process phase $\theta$, $\hat B_i(T_i)$ is a beam splitter operator for mode $i$ with intensity transmission $T_i$, $\hat{a}_\nu$ is the field operator for the vacuum mode $\nu$ that couples through the unused port of the beam splitter, $\hat D_{p}(\alpha)$ and $\hat D_{r}(\beta)$ are the displacement operators for mode $p$ and $r$ with a complex amplitude $\alpha$ and $\beta$, respectively, and $\hat\Phi_p(\phi)$ is the phase rotation operator by a phase $\phi$ for the probe mode. For the bTMSS state, we follow the Yuen notation~\cite{Yuen1976} of squeezing after displacement, $\hat S_{p,r}\hat D_p(\alpha)\hat D_r(\beta)\ket{0,0}$, as this is the typical order of operations when generating these states experimentally. With this notation, $|\alpha|^2$ and $|\beta|^2$ correspond to the number of photons used to seed the parametric process that generates the TMSS.

With these definitions and transformations, the covariance matrix for the state after interaction with the sensor and all external losses can be show to be of the form
\be
\boldsymbol\sigma=\left( \begin{array}{cccc}
                T_p C_{2s}+1-T_p & 0 & 0 & -\sqrt{T_pT_r}e^{i(\theta+\phi)} S_{2s}\\
                0 & T_r C_{2s}+1-T_r & -\sqrt{T_pT_r}e^{i(\theta+\phi)} S_{2s} & 0\\
                0 & -\sqrt{T_pT_r}e^{-i(\theta+\phi)} S_{2s} & T_p C_{2s}+1-T_p & 0\\
                -\sqrt{T_pT_r}e^{-i(\theta+\phi)} S_{2s} & 0 & 0 & T_r C_{2s}+1-T_r
                \end{array}\right),\label{eqn:QCRBderivesigma}
\ee
where $C_{2s}=\cosh(2s)$ and $S_{2s}=\sinh(2s)$. As the transmission in  each mode decreases, the covariance matrix of the bTMSS also tends towards that of a coherent state, which has a covariance matrix of the form $\textrm{diag}(1,1,1,1)$. The displacement vector takes the form
\be
\vec{d}=  \left(\begin{array}{c}
            \sqrt{T_p}\left[\alpha\cosh(s)-\beta^*e^{i\theta}\sinh(s)\right]e^{i\phi}\\
            \sqrt{T_r}\left[\beta\cosh(s)-\alpha^*e^{i\theta}\sinh(s)\right]\\
            \sqrt{T_p}\left[\alpha^*\cosh(s)-\beta e^{-i\theta}\sinh(s)\right]e^{-i\phi}\\
            \sqrt{T_r}\left[\beta^*\cosh(s)-\alpha e^{-i\theta}\sinh(s)\right]\\
            \end{array}\right).\label{eqn:QCRBderivedisp}
\ee
As expected, a decrease in transmission reduces the displacement in phase space and for the limiting case of zero transmission the displacement goes to zero. Additionally, only the probe beam experiences a phase rotation $\phi$.

If we substitute equations~\eqref{eqn:QCRBderivesigma} and~\eqref{eqn:QCRBderivedisp} into equation~\eqref{eqn:QCRBderive} we can show that the QCRB for phase estimation takes the form
\be
\var{\phi} \ge\frac{1}{4T_p\expected{\hat n_p}_0}-\frac{1}{4\expected{\hat n_p}_0}\frac{(2T_r-1)\big[1+2\sinh^2(s)\big]}{1+2T_r\sinh^2(s)}[1-\sech(2s)],
\ee
where $\expected{\hat n_p}_0=|\alpha|^2\cosh^2(s)+|\beta|^2\sinh^2(s)-|\alpha||\beta|\cos(\theta-\chi-\xi)\sinh(2s)$ is the number of probe photons generated. In arriving to this result, we have defined the phases of the coherent states used to seed the squeezing process as $\chi=\arg(\alpha)$ and $\xi=\arg(\beta)$ for the probe and reference, respectively, and have assumed that these phases are known and that there is a phase reference, typically the LO in HD, with which they are compared.
Finally, if we set $T_p=\eta_{p1}T\eta_{p2}$, $T_r=\eta_r$ and $N=\eta_{p1}\expected{\hat n_p}_0$ we arrive at
\be
\var{\phi}_Q^\textrm{TMSS}\ge\frac{1}{4T\eta_{p2}N}-\frac{1}{4N}\eta_{p1} \frac{(2\eta_r-1)\big[1+2\sinh^2(s)\big]}{1+2\eta_r\sinh^2(s)} \left[1-\sech(2s)\right],
\ee
which corresponds to equation~\eqref{EQ:QCRB_TMSS_Phase}.

\section{Phase Transfer Function for Lorentzian and Butterworth Lineshapes}\label{App:Transfer}

In order to compare the transmission- and phase-based sensing schemes, it is necessary to determine the transmission and phase transfer functions for arbitrary on- and off-resonance transmissions. We start with the case of a Lorentzian lineshape, for which it is possible to obtain an analytical expression.  To do so, we must first find the complex amplitude transfer function for the resonance sensor.  The Lorentzian lineshape has a well known complex amplitude transfer function of the form
\be
t_0(\Lambda)=\frac{1}{1-i\Lambda},
\ee
which satisfies the Kramers-Kronig relations.  We can generalize this equation to obtain the transfer function for a resonance sensor with a Lorentzian lineshape and arbitrary $T_\textrm{res}$ and $T_\textrm{off}$ by adding two Lorentzian lineshapes with different amplitudes and linewidths, that is
\be
t(\Lambda)=at_0(\Lambda)+bt_0(\Lambda/\sigma),\label{eqn:generalLorentztransf}
\ee
and take the limit $\sigma\rightarrow\infty$.  This form corresponds to a Lorentzian of arbitrary height, positive or negative, sitting on top of a second extremely wide Lorentzian that gives the off-resonance transmission. Due to the linearity of the Hilbert transform involved in the Kramers-Kronig relations, this general form is a valid physical one. To find the required values of $a$ and $b$ of the amplitude transfer function defined in equation~\eqref{eqn:generalLorentztransf}, we solve for
\be
\left|t(\Lambda)\right|^2=T_\textrm{off}+\frac{T_\textrm{res}-T_\textrm{off}}{1+\Lambda^2},
\ee
which gives
\bea
a&=&\sqrt{T_\textrm{res}}-\sqrt{T_\textrm{off}},\\
b&=&\sqrt{T_\textrm{off}}.
\eea
The phase transfer function is then given by the arctan of the ratio between the imaginary and real parts of the complex amplitude transfer function~\eqref{eqn:generalLorentztransf}, that is
\be
\phi(\Lambda)=\arctan\left[\frac{\Lambda(\sqrt{T_\textrm{res}}-\sqrt{T_\textrm{off}})}{\sqrt{T_\textrm{res}}+\Lambda^2\sqrt{T_\textrm{off}}}\right],
\ee
which corresponds to equation~\eqref{EQ:PhaseLorentz}.

It is not possible to generalize the above approach to obtain an analytical solution for the Butterworth filter of order $m=3$ that we consider in the paper. Therefore, we numerically solve for the minimum phase~\cite{Bechhoefer2011} using the Bode gain-phase relation~\cite{Bodebook}, where the phase is given by the Hilbert transform of the natural log of the magnitude of the amplitude transfer function. We numerically evaluated the required Hilbert transform with two different approaches to validate the results. The first one is based on the fast Fourier transforms (FFT)~\cite{Jerome2012}. This method has a problem with spectral leakage, which causes errors near the extremes of the range used for the FFT. To minimize this problem, we perform the FFT over a generalized wavelength range $\Lambda=\pm1000$ and only use the range $\Lambda=\pm3$ for which the error is negligible. The optimal generalized wavelength for the estimation of the resonance shift  is always inside this range. The second approach, based on Ref.~\cite{Lee1997}, consists in evaluating the phase through the integral
\be
\phi(\Lambda)^{(3)}_\textrm{min}=\frac{1}{\pi}\int_0^\infty\ln\left|\frac{L+\Lambda}{L-\Lambda}\right|\cdot\frac{3L^5\left(T_\textrm{off}-T_\textrm{res}\right)}{\left[1+L^6\right]\left[T_\textrm{off}L^6+T_\textrm{res}\right]}dL,
\ee
where the natural log term is the kernel and the term outside of the natural log is the derivative, with respect to wavelength, of the natural log of the transmission spectrum of the third order Butterworth filter. While this second approach leads to fluctuations in the calculated phase, with sufficient smoothing, it gave the same results as the FFT-based one and served as a validation.

\section{Saturation of QCRB with Optimized Homodyne Detection}\label{App:HD}

In this appendix, we show that an optimized HD can saturate the QCRB for the estimation of both transmission and phase even in the presence of losses external to the resonance sensor.  We start by defining the generalized quadrature operator
\be
\hat Q(\gamma)=\hat ae^{-i\gamma}+\hat a^\dagger e^{i\gamma},
\ee
such that for a coherent state
\bea
\expected{\hat Q(\gamma)}_\textrm{coherent}=2|\alpha|\cos(\chi-\gamma)&\qquad {\rm and}&\qquad
\var{\hat Q(\gamma)}_\textrm{coherent}=1,
\eea
where $\gamma$ determines the quadrature (amplitude or phase), $|\alpha|$ is the displacement of the coherent state, and $\chi$ is the phase of the displacement.

As we show, the optimized HD that saturates the bTMSS QCRB for phase and transmission estimation takes the form $Q_p(\gamma_p)-g^{\rm opt}\hat Q_r(\gamma_r)$, where the subindeces $p$ and $r$ indicate probe or reference mode, respectively, and $g^{\rm opt}$ is the electronic gain on the reference quadrature detection optimized to minimize the variance of the measurement. To calculate the variance of this measurement, we use the quadrature transformations analogous to the ones for the field operators given by equations~\eqref{eqn:Transf1}-\eqref{eqn:Transf6}. We can then show that variance of the optimized HD takes the form
\be
\var{\left[\hat Q_p(\gamma_p)-g^\textrm{opt}\hat Q_r(\gamma_r)\right]}=T_p\cosh(2s)+1-T_p-\frac{T_pT_r\sinh^2(2s)\cos^2(\gamma_p+\gamma_r-\phi-\theta)}{T_r\cosh(2s)+1-T_r},
\ee
where $g^\textrm{opt}=-\frac{\sqrt{T_pT_r}\sinh(2s)\cos(\gamma_p+\gamma_r-\phi-\theta)}{T_r\cosh(2s)+1-T_r}$.  Note that the sign of $g^\textrm{opt}$ changes when considering conjugate quadratures, $\gamma_p\rightarrow\gamma_p+\pi/2$ and $\gamma_r\rightarrow\gamma_r+\pi/2$, consistent with having, for example, reduced noise in the amplitued difference and phase sum of a bTMSS. The noise in the optimized HD is minimized when $\gamma_p+\gamma_r-\phi-\theta=n\pi$ where $n\in \mathbb{Z}$. In what follows, we set the phase of the parametric process that  generate the bTMSS to $\theta = \chi+\xi$ without loss of generality, as its effect can be compensated through the phases of the seeding modes, $\chi$ and $\xi$. Given the nature of the correlations in bTMSSs, we only consider the cases in
which the same quadrature is detected for the probe and reference beams, that is $\gamma_p=\gamma_r\equiv\gamma$.  In this case, the variance of the optimized HD is minimized when $\gamma=(n\pi-\phi-\theta)/2$, with $n$ even, $\gamma_e$, corresponding to the amplitude quadrature and $n$ odd, $\gamma_o$, to the phase quadrature.

We can now determine the variance in the estimation of transmission or phase through error propagation, such that
\be
\var{X}=\frac{\var{\left[\hat Q_p(\gamma)-g^\textrm{opt}\hat Q_r(\gamma)\right]}}{\left|\frac{\partial\expected{\hat Q_p(\gamma)}}{\partial X}\right|^2},
\ee
where $X$ represents transmission or phase. Note that we only take the derivative of the probe quadrature into account as it is the only term that explicitly depends on $X$ given that only the probe beam interacts with the resonance sensor.  Taking into account the mean value of the probe quadrature
\be
\expected{\hat Q_p(\gamma)}=2\sqrt{T_p}\left[|\alpha|\cosh(s)\cos(\chi-\gamma+\phi)-|\beta|\sinh(s)\cos(\xi+\gamma-\theta-\phi)\right],
\ee
setting $T_p=\eta_{p1}T\eta_{p2}$ and $T_r=\eta_r$, and recalling that the mean number of photons interacting with the sensor is
\be
N=\eta_{p1}\left[|\alpha|^2\cosh^2(s) + |\beta|^2\sinh^2(s) - |\alpha||\beta|\cos(\theta-\chi-\xi)\sinh(2s)\right],
\ee
we can show that when $\xi=\chi+\phi$
\bea
\var{T}&=&\frac{\var{\left[\hat Q_p(\gamma_e)-g^\textrm{opt}\hat Q_r(\gamma_e)\right]}}{\left|\frac{\partial\expected{\hat Q_p(\gamma_e)}}{\partial T}\right|^2}\\
&=&\frac{T}{\eta_{p2}N}-\frac{T^2}{N}\eta_{p1} D_r [1-\sech(2s)]\label{eqn:HD_transEstimation}
\eea
and
\bea
\var{\phi}&=&\frac{\var{\left[\hat Q_p(\gamma_o)-g^\textrm{opt}\hat Q_r(\gamma_o)\right]}}{\left|\frac{\partial\expected{\hat Q_p(\gamma_o)}}{\partial \phi}\right|^2}\\
&=&\frac{1}{4T\eta_{p2}N}-\frac{1}{4N}\eta_{p1} D_r [1-\sech(2s)]\label{eqn:HD_phaseEstimation}.
\eea
As we can see, optimized HD always saturates the QCRB when all phases are properly set, which is common among approaches to saturate the QCRB for phase estimation. In practice, one can introduce an extra phase element on the probe beam before the HD to minimized the uncertainty in the estimation of  phase or transmission.


\providecommand{\newblock}{}

\end{document}